\documentclass{webofc}
\usepackage[varg]{txfonts}   
\newcommand{\eqbad}{\stackrel{\textrm{?}}{=}}
\woctitle{POETIC6}
\begin{document}
\title{Non-dipolar gauge links for  transverse-momentum-dependent pion wave functions}
%
%

\author{Yu-Ming Wang \inst{1}\fnsep\thanks{\email{yu-ming.wang@univie.ac.at}}
}

\institute{Fakult\"{a}t f\"{u}r Physik, Universit\"{a}t Wien, Boltzmanngasse 5, 1090 Vienna, Austria
}

\abstract{%
I discuss the factorization-compatible definitions of transverse-momentum-dependent (TMD) pion wave functions
which are fundamental theory inputs entering QCD factorization formulae for many hard exclusive processes.
I will first demonstrate that the soft subtraction factor introduced to remove both rapidity and pinch singularities
can be greatly reduced by making the maximal use of the freedom to construct the Wilson-line paths  when defining
the TMD wave functions. I will then turn to show that the newly proposed TMD definition with non-dipolar Wilson lines
is equivalent to the one with dipolar gauge links and with a complicated soft function, to all orders of
the perturbative expansion in the strong coupling, as far as the infrared behavior is concerned.
}
\maketitle
\section{Introduction}
\label{intro}

Hard exclusive processes have significantly sharpened our understanding of the strong interaction dynamics involved
in the hadronic amplitudes, due to the establishment of various QCD factorization theorems which allow for separation
of the hard scattering contributions and the (universal) long-distance QCD effects.
Albeit the fact that factorization in a non-abelian gauge theory like QCD is by no means a trivial issue,
collinear factorization for a number of interesting hard reactions, at leading power in $1/Q$ where $Q$
is the typical hard scale for a given process, can be  established with rigorous power counting schemes and
standard assumptions. By contrast, transverse-momentum-dependent (TMD) factorization for hard exclusive processes
based upon the theory of on-shell Sudakov form factor \cite{Botts:1989kf}, with the original aim of
avoiding the rapidity divergences in collinear factorization for the pion electromagnetic form factor
at sub-leading power \cite{Li:1992nu}, turns out to be far more sophisticated conceptually due to the absence of
a definite power counting scheme and to the emergence of light-cone singularities in the infrared subtraction,
and this can be already partially  recognized from the troubled history of TMD wave functions
\cite{Lepage:1980fj,Collins:2003fm,Nandi:2007qx,Feng:2008zs,Li:2010nn,Li:2012nk,Wang:2012ab,
Li:2012md,Li:2013xna,Cheng:2014fwa,Shen:2014wga,Wang:2015vqa}.

The primary difficulty of defining TMD wave functions lies in eliminating the rapidity divergence involved in the
infrared subtraction, due to the TMD hard scattering kernels, without introducing additional pinch singularities.
It is the subjective of this presentation to overview the current status of TMD factorization for hard exclusive
processes, focusing on the comparison between collinear and TMD pion wave functions and the regularization of
unwanted rapidity and pinch divergences. I will consider the simplest hard exclusive process $\gamma^{\ast} \to \pi \gamma$
(neglecting the involved sub-leading contributions \cite{Agaev:2010aq} for the moment) through the discussion in order not to
bother about the complication due to the Glauber/Coulomb gluon exchange \cite{Collins:2011zzd} which are relevant to the
construction of factorization theorems for  more complicated processes, e.g., the pion electromagnetic form factor.
It is certainly interesting to generalize the discussion presented here to understand the factorization properties
of those processes taking into account the contribution from the  Glauber gluons, which is also crucial to construct the
Wilson line structures of TMD wave functions. In what follows, I will first explain the origin of light-cone singularities
in the infrared subtraction determined by the convolution of TMD wave functions and the hard functions, and present new
definitions of the TMD pion wave functions with dipolar and non-dipolar Wilson lines in section \ref{section: definition}.
Demonstration of the equivalence of different TMD definitions is then discussed with the aid of rapidity evolution
equations in section \ref{section: equivalence}.

\section{Factorization-compatible TMD wave functions}
\label{section: definition}

To facilitate the comparison between the collinear and TMD pion wave functions,
I will begin with the collinear factorization formula for the pion-photon transition form factor \cite{Lepage:1980fj}
\begin{eqnarray}
F_{\pi}(Q^2)=\frac{\sqrt{2} \,f_{\pi}} {3} \, \int_0^1 dx \, H(x, Q^2,\mu) \, \phi_{\pi}(x, \mu)\,,
\end{eqnarray}
where the pion light-cone distribution amplitude (LCDA)  $\phi_{\pi}(x, \mu)$ is given by
\begin{eqnarray}
\langle 0 |\bar q (0) \, W_{n_{-}}(0, t\, n_{-}) \, \!  \not n_{-} \, \gamma_5 \, q (t \, n_{-}) |  \pi^{+}(p)\rangle
= i \, f_{\pi}\, p_{+} \, \int_0^1 dx \, e^{-i x \, t \, p_{+}} \,\phi_{\pi}(x, \mu) \,,
\label{def: collinear pion DA}
\end{eqnarray}
with the collinear gauge link
\begin{eqnarray}
W_{n_{-}}(0, t\, n_{-}) =P \exp \left[ i \, g_s \int_0^{t} \, d \lambda \, T^a \,
n_{-} \cdot A^a \left( \lambda \, n_{-} \right ) \right] \,.
\end{eqnarray}
The hard function $ H(x, Q^2,\mu)$ determined from the  perturbative  QCD matching is known to the next-to-next-to-leading
order in the strong coupling \cite{Melic:2002ij}. Taking into account the Sudakov form factor to suppress the
 soft gluon contribution \cite{Li:1992nu}, the TMD factorization formula for the $\gamma^{\ast} \to \pi \gamma$ form factor
can be written as
\begin{eqnarray}
F_{\pi}(Q^2)=\frac{\sqrt{2} \,f_{\pi}} {3} \, \int_0^1 dx \,
\int d^2 \vec{b}\,  H(x,  \vec{b}, Q^2,\mu) \, \phi_{\pi}(x,  \vec{b}, \mu)\,
e^{-S(x,\vec{b},Q)}\,,
\end{eqnarray}
where $S(x,\vec{b},Q)$ is the perturbative Sudakov form factor and the non-perturbative Sudakov evolution
is left out  due to the strong model dependence. Naively,  one may expect the following definition of the TMD
pion wave function in momentum space
\begin{eqnarray}
\phi^{\rm naive}_{\pi}(x,\vec{k}_T,\mu) & \eqbad & \int \frac{d z_{-}}{2 \pi} \int \frac{d^2 z_{T}}{(2 \pi)^2}
\,  e^{i( x p_{+}z_{-} -\vec{k}_T \cdot \vec{z}_T)} \,  \nonumber \\
&& \times \,\, \langle 0| \bar q(0) W^{\dag}_{n_-}(+\infty,0)  \not  \! n_{-} \, \gamma_5 \,
[{\rm tr. \,\, link}] \, W_{n_-}(+\infty,z) \, q(z) |\pi^{+}(p)  \rangle \,,
\label{TMD definition: most naive}
\end{eqnarray}
with the coordinate $z=(0, z_{-}, \vec{z}_T)$, which can be constructed from a straightforward extension of
(\ref{def: collinear pion DA}). Unfortunately,  the above-mentioned definition yields the  light-cone divergence
in the infrared subtraction, which can be readily understood from the one-loop correction to the quark-Wilson-line diagram
\begin{eqnarray}
\phi_{\pi}^{(1)} \otimes H^{(0)} \propto  \int [d l] \frac{1}{[(k+l)^2 + i0)][l_+ + i0][l^2+i0]}\,
 \left [ H^{(0)}(x+{l_+/p_+},\vec{k}_T+\vec{l}_T) - H^{(0)}(x,\vec{k}_T) \right ] \,.
\end{eqnarray}
In contrast to the collinear factorization, this loop integral develops an endpoint singularity  from
the $1/l_+$ singularity in the Eikonal propagator, due to the non-cancellation of two hard functions
at $l_+=0$. Two different definitions were proposed \cite{Collins:2003fm} to regularize such rapidity divergence
either by rotating the gauge links off the light cone or by introducing an additional subtraction factor
\begin{eqnarray}
\phi^{C1}_{\pi}(x,\vec{k}_T,y_u,\mu) & \eqbad & \int \frac{d z_{-}}{2 \pi} \int \frac{d^2 z_{T}}{(2 \pi)^2}
\,  e^{i( x p_{+}z_{-} -\vec{k}_T \cdot \vec{z}_T)} \,  \nonumber \\
&& \times \,\, \langle 0| \bar q(0) W^{\dag}_{ u}(+\infty,0)  \not \! n_{-} \, \gamma_5 \,
[{\rm tr. \,\, link}] \, W_{u}(+\infty,z) \, q(z) |\pi^{+}(p)  \rangle \,,
\label{Collins definition: naive 1}\\
\phi^{C2}_{\pi}(x,\vec{k}_T,y_u,\mu) &\eqbad & \int \frac{d z_{-}}{2 \pi} \int \frac{d^2 z_{T}}{(2 \pi)^2}
\,  e^{i( x p_{+}z_{-} -\vec{k}_T \cdot \vec{z}_T)} \,  \nonumber \\
&& \times  \frac{\langle 0| \bar q(0) W^{\dag}_{n_-}(+\infty,0) \,\,  \! \not  n_{-} \, \gamma_5 \,
[{\rm tr. \,\, link}] \, W_{n_-}(+\infty,z) \, q(z) |\pi^{+}(p)  \rangle}
{ \langle 0 | W^{\dag}_{n_-}(+\infty,0) W_{u}(+\infty,0) \,
[{\rm tr. \,\, link}] \, W_{n_-}(+\infty,z) \, W^{\dag}_{u}(+\infty,z) | 0\rangle } \,,
\label{Collins definition: naive 2}
\end{eqnarray}
where the gauge vector $u=(u_{+}, u_{-}, \vec{0}_T)$ is away from the light cone.

Some years later, Bacchetta et al \cite{Bacchetta:2008xw} found  that the TMD  definitions like
(\ref{Collins definition: naive 1}) and (\ref{Collins definition: naive 2}) will induce the pinch singularity
for a TMD parton density with space-like Wilson lines and hence they cannot be used to construct TMD factorization
formulae for the hard inclusive reactions in practice.
Even worse, the pinch singularity appears in  TMD wave functions for any off-light-cone gauge vector, because the radiative
gluon is not necessary to be on the mass shell \cite{Li:2014xda}. To develop a better understanding of the resulting
pinch singularity, one can compute the Wilson line self-energy graph explicitly
\begin{eqnarray}
\phi_{\pi} \propto \int [d l] \frac{u^2}{[l + i0)][u \cdot l + i0][u \cdot l - i0]} \,
\delta(x^{\prime} - x + {l_+/p_+}) \, \delta^{(2)}(\vec{k}_T^{\prime} - \vec{k}_T + \vec{l}_T)   \,,
\end{eqnarray}
which yields the linear divergence in the length of the Wilson lines in  coordinate space.
One then concludes that {\it off-light-cone dipolar Wilson lines regularize the rapidity divergence
at the price of introducing an unwanted pinch singularity}.

\begin{figure}
\centering
\includegraphics[width=10 cm,clip]{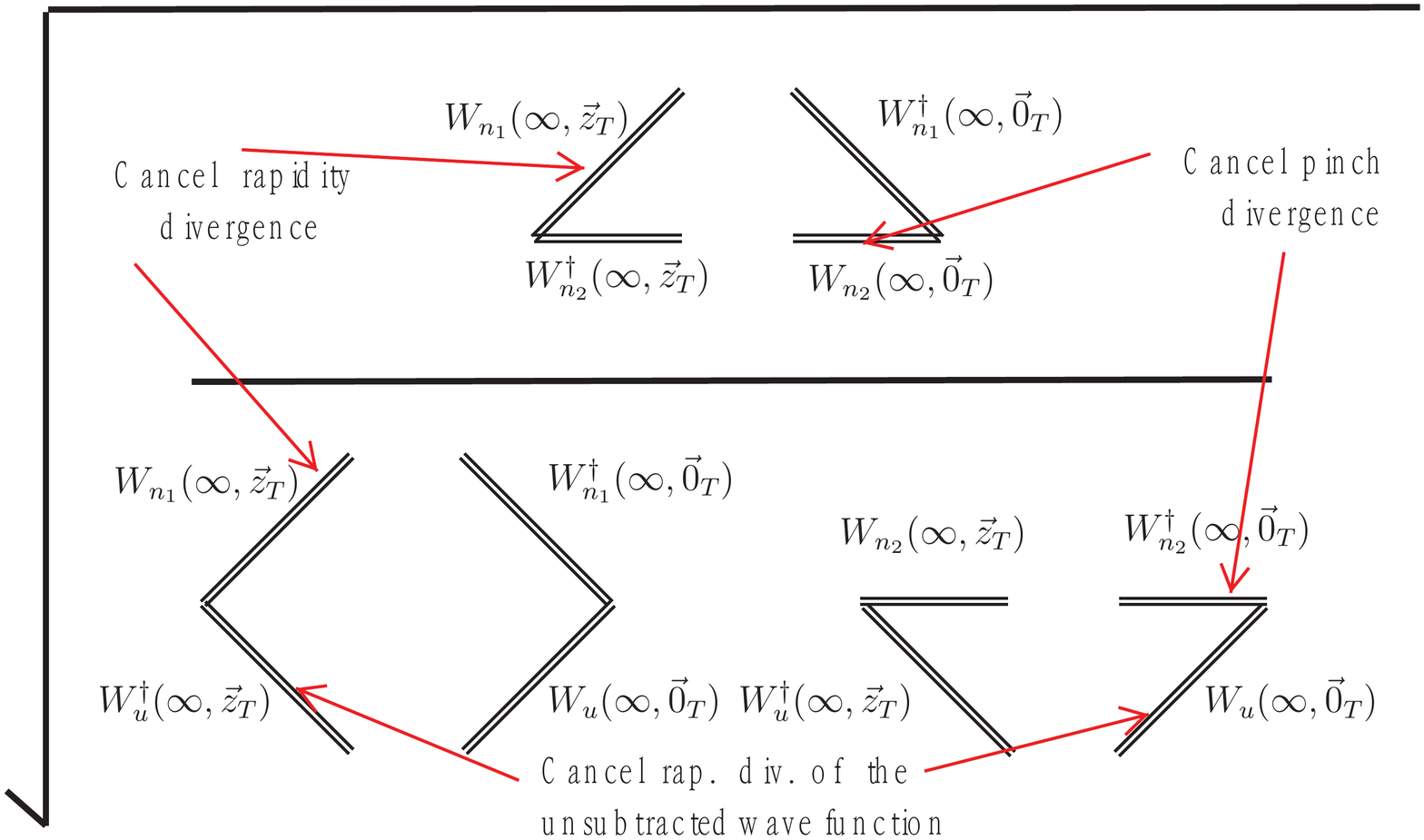}
\caption{Diagrammatical representation of the soft subtraction in (\ref{Collins definition: correct}). }
\label{fig: IR-cancellation}       
\end{figure}

The problem of defining factorization-compatible TMD definitions was resolved only recently by Collins \cite{Collins:2011zzd}
in the context of parton densities. Generalizing his proposal to the TMD wave functions one obtains
\begin{eqnarray}
\phi^{C}_{\pi}(x,\vec{k}_T,y_2,\mu) &=& \lim_{\substack{y_1 \to + \infty  \\   y_u \to - \infty}}  \,\,
\int \frac{d z_{-}}{2 \pi} \int \frac{d^2 z_{T}}{(2 \pi)^2}
\,  e^{i( x p_{+}z_{-} -\vec{k}_T \cdot \vec{z}_T)} \,  \nonumber \\
&& \times \,\,  \langle 0| \bar q(0) W^{\dag}_{u}(+\infty,0)  \not  \! n_{-} \, \gamma_5 \,
[{\rm tr. \,\, link}] \, W_{u}(+\infty,z) \, q(z) |\pi^{+}(p)  \rangle \,  \nonumber \\
&& \times  \,\, \sqrt{\frac{S(z_T; y_1, y_2)} {S(z_T; y_1, y_u) \, S(z_T; y_2, y_u)} } \,,
\label{Collins definition: correct}
\end{eqnarray}
where the soft function is defined as
\begin{eqnarray}
S(z_T;y_A,y_B) =\frac{1}{N_c} \langle 0|W^{\dag}_{n_B}(\infty,\vec{z}_T)_{ca} \,
W_{n_A}(\infty,\vec{z}_T)_{ad} \,  W_{n_B}(\infty,0)_{bc} \, W^{\dag}_{n_A}(\infty,0)_{db} | 0 \rangle,
\end{eqnarray}
with $n_A$ and $n_B$ being the rapidities of the gauge-fixing vectors $n_A$ and $n_B$.
This definition is unique, up to the choice of the ultraviolet  renormalization prescription, provided that
the following criteria are satisfied \cite{Collins:2011ca}.
\begin{itemize}
\item {The unsubtracted wave function only involves light-cone Wilson lines.}
\item {Each soft factor has at most one off-light-cone Wilson line.}
\item {The subtracted wave function is free of rapidity and pinch singularities
and is defined as the product of the unsubtracted wave function and powers of the soft function
in transverse coordinate space. }
\end{itemize}
To demonstrate the cancellation mechanism in a transparent way,  a detailed account of the rapidity
 and pinch structures of different pieces in the soft subtraction is  display in figure \ref{fig: IR-cancellation}.

\begin{figure}
\centering
\includegraphics[width=8 cm,clip]{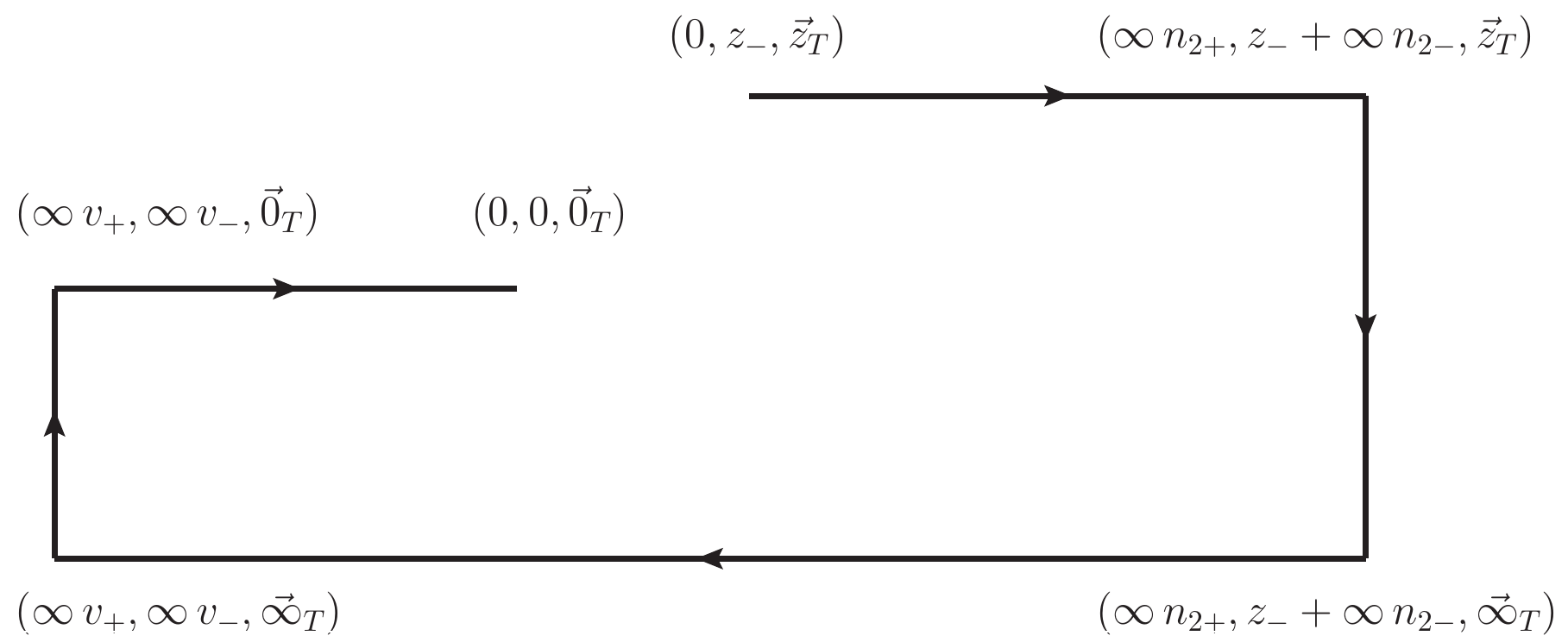}
\caption{The Wilson-line path adopted to define the  unsubtracted wave function
(\ref{New definition: general}). }
\label{fig: Wilaon-line path}       
\end{figure}

At this point,  one might be curious to ask whether there exists an alternative definition of the TMD pion wave function
with a simpler soft subtraction function. The key to answering this question consists in  identifying the very origin
of pinch singularities, which is due to {\it the dipolar structure of the gauge links} defining the TMD wave functions
or parton  densities. It is then a straightforward task to propose the following definition \cite{Li:2014xda}
\begin{eqnarray}
\phi^{\rm new}_{\pi}(x,\vec{k}_T,y_2,\mu) &=&
\int \frac{d z_{-}}{2 \pi} \int \frac{d^2 z_{T}}{(2 \pi)^2}
\,  e^{i( x p_{+}z_{-} -\vec{k}_T \cdot \vec{z}_T)} \,  \nonumber \\
&& \times \,\,  \frac{\langle 0| \bar q(0) W^{\dag}_{n_2}(+\infty,0)  \not \! n_{-} \, \gamma_5 \,
[{\rm links @ \infty}] \, W_{v}(+\infty,z) \, q(z) |\pi^{+}(p)  \rangle}
{[{\rm color \,\, factor}] \,\,\langle 0 | W^{\dag}_{n_2}(+\infty,0) \,
[{\rm links @ \infty}] \, W_{v}(+\infty,0)| 0\rangle} \,,
\label{New definition: general}
\end{eqnarray}
with two {\it distinct} off-light-cone gauge vectors $n_2$ and $v$.
It is evident that the pinch singularity appeared in the naive definition (\ref{Collins definition: naive 1})
is now  alleviated to a soft divergence in the unsubtracted wave function
\begin{eqnarray}
\phi^{\rm new}_{\pi} \supset \int [d l] \frac{n_2 \cdot v}{[l + i0)][n_2 \cdot l + i0][v \cdot l - i0]}\,
\delta(x^{\prime} - x + {l_+/p_+}) \delta^{(2)}(\vec{k}_T^{\prime} - \vec{k}_T + \vec{l}_T)   \,,
\end{eqnarray}
which is further subtracted by  the soft function with a rather simple gauge structure.
The Wilson-line path  defining  the unsubtracted wave function is presented in figure \ref{fig: Wilaon-line path}
and a similar path of Wilson lines at infinity is used to define the soft function.
Different choices for the gauge-link path may serve the same purpose, however, a more careful study is in demand to
quantity the scheme dependence on the Wilson-line contour defining the TMD wave functions (see e.g., \cite{Cherednikov:2014mua}).

The general definition (\ref{New definition: general}) can be further reduced with some
specific choices of two non-light-cone vectors $n_2$ and $v$.
In particular,  employing the orthogonal Wilson lines ($n_2 \cdot v=0$) no soft subtraction
will be needed in the TMD definition \cite{Li:2014xda}
\begin{eqnarray}
\phi^{\rm I}_{\pi}(x,\vec{k}_T,y_2,\mu) &=&
\int \frac{d z_{-}}{2 \pi} \int \frac{d^2 z_{T}}{(2 \pi)^2}
\,  e^{i( x p_{+}z_{-} -\vec{k}_T \cdot \vec{z}_T)} \,  \nonumber \\
&& \times \,\, \langle 0| \bar q(0) W^{\dag}_{n_2}(+\infty,0) \not \!  n_{-} \, \gamma_5 \,
[{\rm links @ \infty}] \, W_{v}(+\infty,z) \, q(z) |\pi^{+}(p)  \rangle \,,
\label{New definition: orthogonal}
\end{eqnarray}
with
\begin{eqnarray}
n_2 =(e^{y_2}, e^{-y_2}, \vec{0}_T)  \,, \qquad v =(-e^{y_2}, e^{-y_2}, \vec{0}_T)\,.
\end{eqnarray}
One can readily verify that both of the two definitions (\ref{Collins definition: correct})
and (\ref{New definition: orthogonal}) result in the same collinear logarithm for the infrared subtraction
(without the contribution from the LSZ terms)
\begin{eqnarray}
\phi_{\pi}^{C, \, \rm I} \otimes H^{(0)}
=-{\alpha_s \, C_F \over 4 \pi} \, \left [  2 \, \ln x + 3 \right ] \,
\ln \left ( {k_T^2 \over Q^2 } \right ) \,\,\, H^{(0)}(x,k_T) + {\rm (IR \,\, finite \,\, terms)} \,,
\label {collinear logs of effective diagrams}
\end{eqnarray}
at  one-loop approximation, which reproduces the collinear behaviour of the one-loop QCD amplitude
for the partonic Green function corresponding to $\gamma^{\ast} \to \pi \gamma$ at leading Fock-state approximation.
Alternatively, one can facilitate the TMD definition (\ref{New definition: general})  with the antiparallel  Wilson lines
\begin{eqnarray}
\phi^{\rm II}_{\pi}(x,\vec{k}_T,y_2,\mu) &=&
\int \frac{d z_{-}}{2 \pi} \int \frac{d^2 z_{T}}{(2 \pi)^2}
\,  e^{i( x p_{+}z_{-} -\vec{k}_T \cdot \vec{z}_T)} \,  \nonumber \\
&& \times \,\, \frac{\langle 0| \bar q(0) W^{\dag}_{n_2}(+\infty,0) \not \!  n_{-} \, \gamma_5 \,
[{\rm links @ \infty}] \, W_{n_2}(-\infty,z) \, q(z) |\pi^{+}(p)  \rangle}
{[{\rm color \,\, factor}] \,\,\langle 0 | W^{\dag}_{n_2}(+\infty,0) \,
[{\rm links @ \infty}] \, W_{n_2}(-\infty,0)| 0\rangle} \,,
\end{eqnarray}
which  can be also shown to reproduce the very collinear structure in (\ref{collinear logs of effective diagrams})
and I will not expand the discussion for this definition.

\section{Equivalence of TMD definitions}
\label{section: equivalence}

Now I will demonstrate the equivalence of different TMD definitions to all orders of the strong coupling
in light of  the arguments inspired from the rapidity evolution equations.
First,  I will explain that both of the definitions (\ref{Collins definition: correct}) and
(\ref{New definition: orthogonal}) will recover the naive definition (\ref{TMD definition: most naive})
in the limit of vanishing regulators. One can readily verify that switching off the regulator in
(\ref{Collins definition: correct}), i.e.,  $ y_2 \to -\infty \left ( \Rightarrow y_2 =y_u \right )$ implies that
\begin{eqnarray}
\sqrt{\frac{S(z_T; y_1, y_2)} {S(z_T;y_1,y_u) \, S(z_T;y_2,y_u)} } \, \rightarrow  1  \,,
\end{eqnarray}
which evidently reduces the Collins' definition to (\ref{TMD definition: most naive}).
Similarly, setting $y_2 \to -\infty \left (\Rightarrow n_2 =v=n_{-} \right )$ in (\ref{New definition: orthogonal})
trivially gives rise to
\begin{eqnarray}
\phi^{\rm I}_{\pi}(x,\vec{k}_T,y_2,\mu) \to \phi^{\rm naive}_{\pi}(x,\vec{k}_T,\mu) \,.
\end{eqnarray}
We are then led to conclude that both Collins' and our proposals for the TMD definitions yield the same
collinear divergence in the limit $y_2=y_u \to -\infty$.
The remaining task is to show that the two definitions can bring about the same collinear behaviour for
an arbitrary rapidity $y_2$. To this end, we need to examine the rapidity evolution equations for both definitions
of the TMD pion wave function, employing the standard resummation techniques in QCD \cite{Collins:1981uk,Botts:1989kf}.

Noticing that the rapidity dependence in Collins' definition (\ref{Collins definition: correct}) comes from the soft
subtraction entirely and applying the routine chain rule for the rapidity derivative one can write
\begin{eqnarray}
\frac{d}{dy_2}\phi_{\pi}^C &=& \lim_{\substack{y_1 \to + \infty  \\   y_u \to - \infty}}  \,
\frac{1}{2}\left[\underbrace{\frac{S^{\prime}(z_T; y_1, y_2)}{S(z_T; y_1, y_2)}}-
\underbrace{\frac{S^{\prime}(z_T;y_2,y_u)}{S(z_T;y_2,y_u)}} \right] \,\, \phi_{\pi}^C
\,\, \approx \lim_{\substack{y_1 \to + \infty}} \,  \ K(z_T; y_1, y_2) \,\, \phi_{\pi}^C  \,, \nonumber \\
&&  \hspace{1.5 cm}  \ K(z_T; y_1, y_2)
\hspace{0.5 cm}  K(z_T; y_2, y_u) \,
\end{eqnarray}
where the primed soft functions $S^{\prime}$ are  computed from the Wilson-line diagrams defining
the soft functions, with the Eikonal vertex $n_2^{\mu}$ replaced by a special vertex given by
Eq. (4.4) in \cite{Li:2014xda}, and only the soft function $K(z_T; y_2, y_u)$  enters the
rapidity evolution of the TMD definition (\ref{Collins definition: correct}) in contrast to
the original Collins-Soper-Sterman evolution \cite{Collins:1981uk,Botts:1989kf}.

\begin{figure}
\centering
\includegraphics[width=6.0 cm,clip]{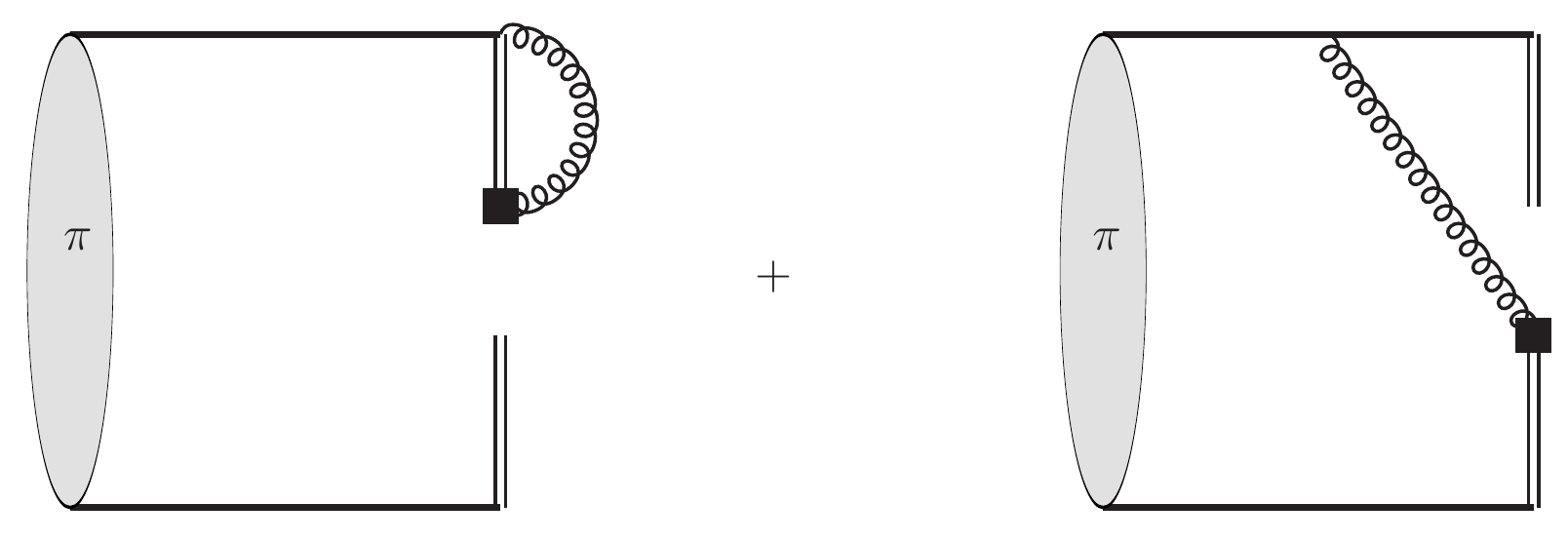} \hspace{1.0 cm}
\includegraphics[width=6.0 cm,clip]{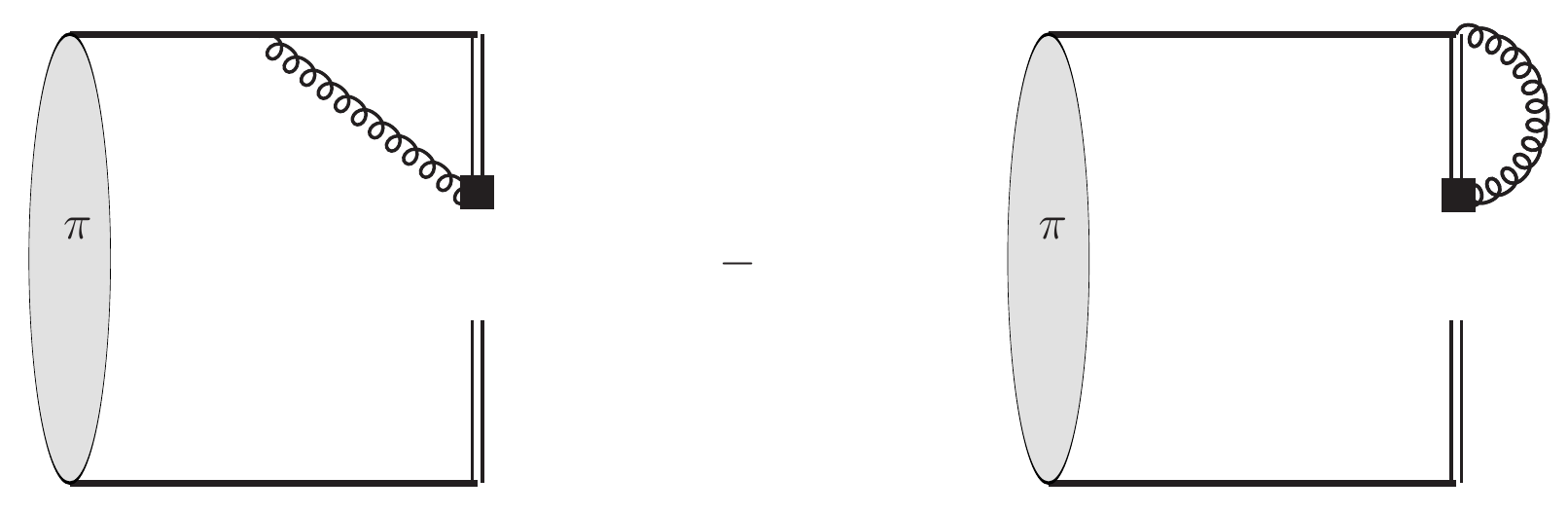} \\
\vspace{0.5 cm}
{\rm K \,\, function} \hspace{5.0 cm} {\rm G \,\, function}
\caption{Graphic representation of the rapidity evolution kernels for
the new TMD definition (\ref{New definition: orthogonal}).}
\label{fig: K and G functiond}       
\end{figure}

Constructing the rapidity evolution for the newly proposed definition (\ref{New definition: orthogonal})
is slightly more involved due to the appearance of two gauge vectors dependent on the same rapidity
parameter $y_2$ and  to the rapidity dependence of the unsubtracted wave function.
It is nevertheless without the conceptual difficulty to derive the following evolution equation
\begin{eqnarray}
\frac{d}{dy_2} \phi^{\rm I}_{\pi} &=& \lim_{y_1 \to + \infty}
\, [K(z_T; y_1, y_2)+G(y_2)] \, \phi^{\rm I}_{\pi}  \,,
\end{eqnarray}
where the soft and hard functions can be computed from the effective diagrams presented in
figure \ref{fig: K and G functiond}. We can then readily conclude that the infrared behaviour of
the TMD definition (\ref{New definition: orthogonal}) is equivalent to that proposed by Collins
(\ref{Collins definition: correct}), to all orders of  the strong coupling, due to the infrared
safety of the hard function  $G(y_2)$.

\section{Concluding discussion}
\label{section: conclusion}

To summarize,  I presented a brief historic report on understanding the TMD factorization for
hard exclusive processes, using the definition of TMD wave functions  as an illustrative example.
I provided a detailed account of justifying the new definition (\ref{New definition: general}),
with a simple soft function, applicable  to a QCD description of $\gamma^{\ast} \to \pi \gamma$
at large momentum transfer. Furthermore, I demonstrated the equivalence of Collins' and our proposals
for the TMD definitions in the infrared behaviours to all orders of perturbative expansion,
with the aid of the rapidity evolution equations  using the classical QCD resummation techniques.
However, there remain  many conceptual issues of constructing  TMD factorization
formulae for hard exclusive reactions as already emphasized above.
Among these unsettled questions, the top priority should be to establish a definite
power counting scheme for all the energy scales involved in a given problem,
including the {\it intrinsic} transverse momentum. I therefore anticipate exciting developments
of putting TMD factorization for hard exclusive processes on a solid ground, in addition to its
phenomenological success.

\section*{Acknowledgements}

I am grateful to Hsiang-nan Li for a very fruitful collaboration and to
the organizers of POETIC6 for the generous finical support.

\end{document}